\documentclass[12pt, a4paper]{article}

\usepackage[utf8]{inputenc}
\usepackage[T1]{fontenc}
\usepackage{graphicx} 
\usepackage{amsmath}
\usepackage{amsfonts}
\usepackage{amssymb}
\usepackage{booktabs} 
\usepackage{caption} 
\usepackage[margin=2.5cm]{geometry} 
\usepackage{hyperref} 
\usepackage{parskip} 
\usepackage{float} 
\usepackage{lscape} 
\usepackage{array} 
\usepackage{longtable} 
\usepackage{cite} 

\hypersetup{
    colorlinks=true,
    linkcolor=blue,
    filecolor=magenta,      
    urlcolor=cyan,
    pdftitle={Advancing Hearing Assessment},
    pdfpagemode=FullScreen,
}

\title{Advancing Hearing Assessment: An ASR-Based Frequency-Specific Speech Test for Diagnosing Presbycusis}
\author{Stefan Bleeck \\ Institute of Sound and Vibration Research, University of Southampton}
\date{\today} 

\begin{document}

\maketitle
\begin{abstract}
Traditional audiometry often fails to fully characterize the functional impact of hearing loss on speech understanding, particularly supra-threshold deficits and frequency-specific perception challenges in conditions like presbycusis. This paper presents the development and simulated evaluation of a novel Automatic Speech Recognition (ASR)-based frequency-specific speech test designed to provide granular diagnostic insights. Our approach leverages ASR to simulate the perceptual effects of moderate sloping hearing loss by processing speech stimuli under controlled acoustic degradation and subsequently analyzing phoneme-level confusion patterns. Key findings indicate that simulated hearing loss introduces specific phoneme confusions, predominantly affecting high-frequency consonants (e.g., alveolar/palatal to labiodental substitutions) and leading to significant phoneme deletions, consistent with the acoustic cues degraded in presbycusis. A test battery curated from these ASR-derived confusions demonstrated diagnostic value, effectively differentiating between simulated normal-hearing and hearing-impaired listeners in a comprehensive simulation. This ASR-driven methodology offers a promising avenue for developing objective, granular, and frequency-specific hearing assessment tools that complement traditional audiometry. Future work will focus on validating these findings with human participants and exploring the integration of advanced AI models for enhanced diagnostic precision.
\end{abstract}

\section{Introduction}
Traditional audiometry, while fundamental for quantifying hearing loss, often provides an incomplete picture of an individual's ability to understand speech, especially in challenging environments. A significant limitation is the poor correlation between pure-tone audiogram (PTA) results and self-reported hearing difficulties or performance on speech perception tests. This discrepancy is particularly evident in supra-threshold deficits, which involve distortions in sound processing above the hearing threshold, such as reduced frequency resolution, impaired temporal processing, and loudness recruitment. These issues are not captured by simple audibility measures.

It is crucial to recognise that for mild-to-moderate hearing loss, a substantial part of the experienced difficulty, particularly in quiet or low-noise conditions, can often be attributed primarily to audibility. This refers to the simple inability to detect certain speech sounds, especially low-intensity, high-frequency consonants (e.g., /s/, /f/, /t/) which are critical for speech clarity \cite{Plomp1978, Humes2007}. This 'attenuation' aspect, as described by Plomp, frequently represents the initial barrier to communication \cite{Plomp1978}. However, supra-threshold distortions—such as reduced frequency selectivity, impaired temporal processing, and altered neural coding—are also inherent consequences of sensorineural hearing loss and play a significant role \cite{Oxenham2008, Henry2022}. While audibility may be the dominant factor in quiet and for mild losses, the impact of supra-threshold deficits becomes increasingly critical as background noise increases and as hearing loss progresses into the moderate range. In these challenging conditions, even when speech is amplified to be technically audible, these distortions can become the primary limiting factor preventing clear speech understanding \cite{Henry2022}. Nonetheless, a strong, predictable link exists between specific patterns of frequency-specific audibility loss and the resulting phonetic confusions \cite{Giguere1995}. This provides a compelling rationale for developing tests that can map these audibility-driven confusions, complementing traditional audiometry by offering valuable insights into an individual's specific perceptual challenges.

Recent advancements in Automatic Speech Recognition (ASR) systems, particularly those built on deep neural networks, offer a promising avenue to overcome these limitations. ASR can automate speech audiometry with accuracy and reliability comparable to human manual scoring, significantly reducing clinician time \cite{mdpi2025, meyer2015}. Crucially, ASR facilitates phoneme-level scoring, providing more detailed data on perception errors by identifying specific phoneme confusions rather than just whole-word misses \cite{hnath2016,fontan_nd1}. This capability allows for a more fine-grained, frequency-specific diagnostic insight into speech perception deficits. This paper details the development of a novel ASR-based frequency-specific speech test designed to provide such insights, particularly for conditions like presbycusis. Our approach leverages ASR to simulate human listeners under controlled acoustic degradation, mimicking typical hearing loss conditions. By meticulously simulating typical hearing loss conditions and analyzing the resulting ASR confusion patterns at the phonetic level, we aim to create a test capable of highlighting specific frequency regions impacted by perceptual degradation, offering a "confusion profile" that complements traditional audiometric data.

\subsection*{Research Questions}
\begin{itemize}
    \item How can ASR be effectively utilized to simulate the perceptual effects of hearing loss and generate comprehensive phoneme confusion profiles?
    \item What specific phoneme-level confusion patterns emerge under simulated moderate sloping hearing loss, and how do these patterns correlate with the acoustic properties of speech sounds and known effects of presbycusis?
    \item Can a test battery curated from these ASR-derived confusions effectively differentiate between normal-hearing and hearing-impaired listeners in a simulated environment?
    \item What are the critical considerations for translating such an ASR-based diagnostic test into a practical and diagnostically valuable tool for human participants?
\end{itemize}

\section{Methodology}
The development of the ASR-based frequency-specific speech test followed a modular and iterative methodology, primarily implemented through custom software modules developed in a scientific computing environment. This encompassed speech material preparation, ASR processing, hearing loss simulation, phoneme analysis, test item curation, and diagnostic simulation.
\subsection{Speech Material and ASR Baseline}
The TIMIT Acoustic-Phonetic Continuous Speech Corpus was selected for its high audio quality, diverse speaker set, and rich phonetic and word-level transcriptions. Individual word audio files were extracted from TIMIT sentences. To establish a robust baseline, a pre-trained ASR model (e.g., based on wav2vec 2.0 architecture) was applied to all extracted clean (unfiltered) words. Raw ASR outputs were lexically normalized using Levenshtein distance against a comprehensive vocabulary to identify the closest valid English word. A phonetic lexicon (e.g., derived from CMUdict) served as the primary phonetic reference for the project, used for obtaining phonetic transcriptions of ASR outputs and for cleaning raw ASR transcriptions. A reverse lookup mechanism was also constructed from this lexicon to facilitate the generation of phonetically plausible distractor words.
\subsection{Hearing Loss Simulation and Phoneme Confusion Analysis}
Sensorineural hearing loss, particularly presbycusis, was simulated using digital filtering. A custom function was developed to apply frequency-dependent attenuation via a Finite Impulse Response (FIR) filter based on user-defined audiogram profiles. Representative audiogram profiles were defined: Normal Hearing (0 dB HL), Mild Hearing Loss (5 dB HL at 250 Hz to 50 dB HL at 8000 Hz), and Moderate Hearing Loss (10 dB HL at 250 Hz to 70 dB HL at 8000 Hz). Pink noise was added to the speech stimuli at a configurable Signal-to-Noise Ratio (SNR), typically 10 dB, with appropriate lead-in and ramp-up/down phases. The ASR system then processed the degraded speech, and its outputs were lexically normalized. For phoneme analysis, ground truth phonetic transcriptions for each original word were extracted. For the cleaned ASR transcription, phonetic transcriptions were obtained by querying the phonetic lexicon. A phoneme-level analysis script combined these to calculate Phoneme Error Rate (PER) using a custom Levenshtein distance algorithm for phoneme sequences, quantifying substitutions, insertions, and deletions. A phoneme-level confusion matrix was generated, counting the frequency of original-to-transcribed phoneme mappings. This analysis involved a backtracing mechanism to identify the specific sequence of operations (Match, Substitution, Deletion, Insertion), with a focus on identifying targeted confusions in high-frequency and mid-to-high frequency ranges.
\subsection{Test Item Curation}
The objective was to curate a set of 200 diagnostically targeted word pairs based on detailed phoneme confusion data. Input data sources included detailed phoneme confusion examples (comparing phoneme sequences recognized by ASR under clean and simulated HL conditions) and the reverse phonetic lexicon. A two-phase test item selection strategy was employed: Phase 1 prioritized the top 10 most frequent non-match confusion types, selecting distinct word pairs exemplifying these errors. Phase 2 proportionally filled the remaining slots, aiming for an overall error distribution mirroring observed ASR errors (approx. 52.7\% Substitution, 34.9\% Deletion, 12.4\% Insertion). General filtering criteria for candidate pairs included a word Levenshtein distance of $\leq$2 between target and distractor, exact syllable count matching, phoneme validity in the lexicon, and a maximum phoneme Levenshtein distance of 8. Distractor words were prioritized from actual ASR outputs under HL or generated phonetically. Each item was categorized by the frequency relevance of the involved clean phoneme (e.g., 'High', 'Mid-High') to align with characteristics of presbycusis.
\subsection{Test Administration and Diagnostic Simulation}
A modular test script was designed for two primary modes: a human test mode (playing degraded audio with a graphical user interface for user input) and an ASR simulation mode (simulating a listener's response using the ASR model, with probabilistic choice determined by Levenshtein distance to the two options). All audio stimuli underwent RMS normalization to a consistent level, optional pink noise addition, and optional listener hearing loss simulation. A comprehensive simulation framework was developed to evaluate the diagnostic capability of the ASR-based test. The primary objective was to quantify its ability to differentiate between simulated normal-hearing (NH) and hearing-impaired (HI) listeners by analyzing Receiver Operating Characteristic (ROC) curves, sensitivity, and specificity. A configurable number of NH and HI participants were simulated, with each participant's performance being independent. HI participants received unique audiogram profiles representing moderate sloping hearing loss, consistent with presbycusis, with random perturbations of $\pm$10 dB around a base audiogram. Diagnostic performance was assessed across varying test lengths (50, 100, and 200 word pairs), with each simulated participant receiving a unique, random subset. The ASR's "choice" was determined by the smaller phonetic Levenshtein distance between its output and the original/distractor words. The overall percent correct score was calculated for each participant.
\subsection{Performance Evaluation and Diagnostic Item Selection}
Mean and standard deviation of scores were calculated for NH and HI groups. ROC analysis was performed using standard statistical functions. NH participants were designated as the 'positive' class, and HI participants as the 'negative' class, with higher scores indicating normal hearing. Area Under the Curve (AUC) was calculated as a scalar measure of discriminative power (0.5 to 1.0). Youden's J statistic (max(Sensitivity+Specificity-1)) identified an optimal operating threshold. Sensitivity and specificity were calculated at this optimal threshold from the perspective of detecting hearing impairment. To refine the speech test for optimal diagnostic capability, an ASR-based diagnostic value assessment was performed. This involved comparing ASR performance under two distinct conditions: Normal Hearing (NH) Simulation (ASR processing speech stimuli with noise but without simulated hearing loss) and Hearing Loss (HL) Simulation (ASR processing the same speech stimuli with noise and simulated moderate sloping sensorineural hearing loss). For each word pair, across multiple Signal-to-Noise Ratio (SNR) levels (e.g., 5 dB, 10 dB, 20 dB), the ASR's correct recognition percentage was recorded for both NH and HL simulations. A "Diagnostic Difference" was calculated as:
\[ \text{Diagnostic Difference} = \text{NH Correct Percentage} - \text{HL Correct Percentage} \]
A higher positive Diagnostic Difference indicated greater diagnostic value. The SNR level maximizing this difference was identified as optimal for differentiation. Only word pairs exceeding a predefined diagnostic difference threshold (e.g., 5\%) were included in the final test battery. Finally, a human speech test implementation was configured to load these diagnostic word pairs and randomize their presentation. It generated acoustic stimuli by mixing original words with noise and optionally filtering for simulated hearing loss. A compact, centered graphical user interface presented two interactive buttons for user input. Post-test, human participant scores were compared to the mean correct percentages of the NH and HL ASR models at the same SNR. The user was then assessed as being in the "Normal Hearing" or "Hearing Loss" category based on which ASR model's performance their score was closer to, providing a preliminary, model-based indication of their hearing status.

\section{Results}
This section presents the key findings from the ASR-based phoneme confusion analysis, test item curation, and diagnostic simulation.
\subsection{Phoneme Confusion Analysis}
\begin{figure}[H]
    \centering
    \includegraphics[width=0.8\textwidth]{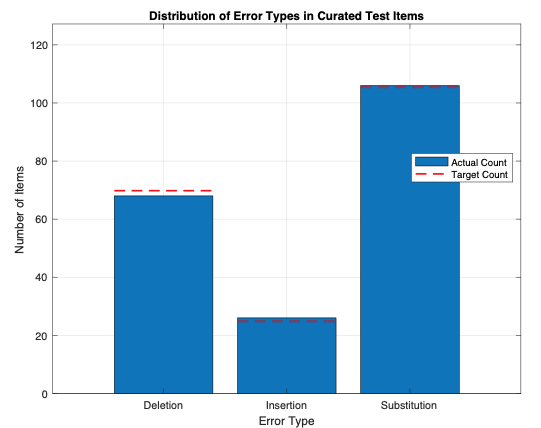} 
    \caption{Distribution of Error Types in Curated Test Items. This bar chart illustrates the distribution of error types (Substitution, Deletion, Insertion) in the final curated test item set (N=200). Blue bars represent the actual number of items selected for each error type, while dashed red lines indicate target counts derived from the overall error type percentages observed in the comprehensive ASR confusion dataset (Substitution: 52.7\%, Deletion: 34.9\%, Insertion: 12.4\%). This figure demonstrates the effectiveness of the two-phase curation strategy in achieving a representative balance of error mechanisms.}
    \label{fig:error_distribution}
\end{figure}

The comprehensive phoneme-level analysis revealed significant degradation patterns in ASR performance under simulated hearing loss, with a total of 29,997 phoneme-level discrepancies observed. The distribution of error types was as follows:
\begin{itemize}
    \item Substitutions: 15,814 occurrences (approx. 52.7\%)
    \item Deletions: 10,459 occurrences (approx. 34.9\%)
    \item Insertions: 3,724 occurrences (approx. 12.4\%)
\end{itemize}
This distribution indicates that substitutions were the most common error, followed closely by deletions, suggesting that the simulated hearing loss frequently caused phonemes to be misidentified or entirely missed by the ASR. Figure \ref{fig:error_distribution} illustrates the distribution of these error types (Substitution, Deletion, Insertion) in the final curated test item set (N=200). The figure compares the actual number of items selected for each error type against target counts derived from the overall error type percentages observed in the comprehensive ASR confusion dataset, demonstrating the effectiveness of the curation strategy in achieving a representative balance of error mechanisms.

The top 20 specific phoneme confusions observed are detailed in Table \ref{tab:phoneme_confusions}.
\begin{table}[H]
    \centering
    \caption{The top 20 specific phoneme confusions observed.}
    \label{tab:phoneme_confusions}
    \begin{tabular}{@{}lc@{}}
        \toprule
        Original $\rightarrow$ Confused & Occurrences \\ \midrule
        S $\rightarrow$ F             & 251         \\
        IY1 $\rightarrow$ EY1         & 212         \\
        S $\rightarrow$ T             & 201         \\
        IH1 $\rightarrow$ EH1         & 200         \\
        W $\rightarrow$ B             & 186         \\
        T $\rightarrow$ D             & 172         \\
        AE1 $\rightarrow$ EH1         & 163         \\
        N $\rightarrow$ L             & 155         \\
        IY1 $\rightarrow$ IH1         & 147         \\
        T $\rightarrow$ K             & 142         \\
        R $\rightarrow$ B             & 140         \\
        IH1 $\rightarrow$ IY1         & 136         \\
        M $\rightarrow$ L             & 134         \\
        S $\rightarrow$ K             & 134         \\
        N $\rightarrow$ T             & 131         \\
        R $\rightarrow$ T             & 129         \\
        OW1 $\rightarrow$ AA1         & 123         \\
        R $\rightarrow$ K             & 122         \\
        AA1 $\rightarrow$ AH1         & 121         \\
        AO1 $\rightarrow$ AA1         & 114         \\ \bottomrule
    \end{tabular}
\end{table}

\begin{figure}[H]
    \centering
    \includegraphics[width=0.8\textwidth]{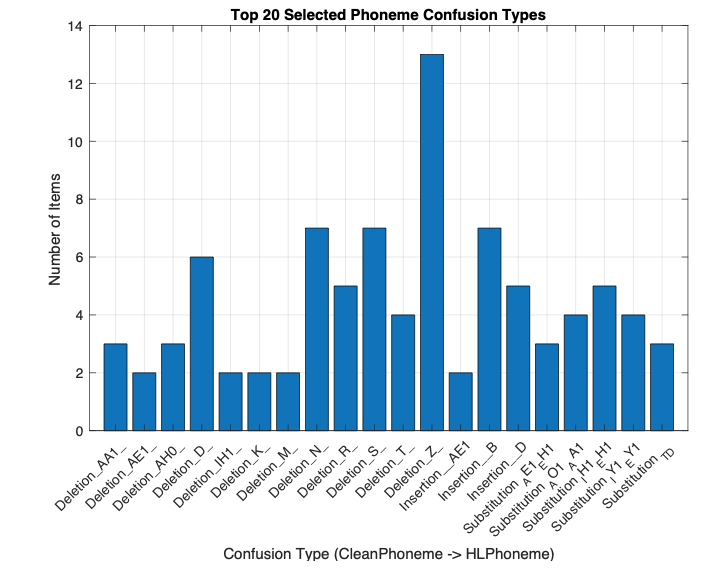} 
    \caption{Top N Selected Phoneme Confusion Types. This bar chart presents the top N (e.g., N=20) most frequently selected specific phoneme confusion types from the curated test item set. Each bar represents a unique confusion key, formatted as 'ErrorType\_CleanPhonemeInvolved\_HLPhonemeInvolved' (e.g., 'Substitution\_S\_F' or 'Deletion\_Z\_'). The height of each bar indicates the number of times that specific confusion type is represented. This figure highlights the individual phoneme-level errors prioritized by the Phase 1 selection, such as common deletions of sibilants or specific vowel substitutions, which are particularly relevant to presbycusis.}
    \label{fig:top_n_confusions}
\end{figure}

Figure \ref{fig:top_n_confusions} presents the top N (e.g., N=20) most frequently selected specific phoneme confusion types from the curated test item set. This highlights the individual phoneme-level errors prioritized by the Phase 1 selection, such as common deletions of sibilants or specific vowel substitutions, which are particularly relevant to presbycusis.

Further analysis of substitution errors based on the place of articulation provided strong support for the hypothesis that high-frequency attenuation caused the observed confusions. The top 10 place of articulation confusions observed from the 15,814 substitution errors are detailed in Table \ref{tab:articulation_confusions}.
\begin{table}[H]
    \centering
    \caption{Top 10 Place of Articulation Confusions (Substitution Errors).}
    \label{tab:articulation_confusions}
    \begin{tabular}{@{}r >{\raggedright\arraybackslash}p{5cm} c@{}}
        \toprule
        Rank & Clean ASR Place $\rightarrow$ HL ASR Place & Count \\ \midrule
        1 & Alveolar/Palatal $\rightarrow$ Labiodental & 251   \\
        2 & Alveolar/Palatal $\rightarrow$ Alveolar/Palatal & 201   \\
        3 & Alveolar/Palatal $\rightarrow$ Bilabial     & 186   \\
        4 & Alveolar/Palatal $\rightarrow$ Velar/Palatal & 142   \\
        5 & Bilabial $\rightarrow$ Alveolar/Palatal    & 140   \\
        6 & Alveolar/Palatal $\rightarrow$ Dental      & 131   \\
        7 & Velar/Palatal $\rightarrow$ Alveolar/Palatal & 122   \\
        8 & Vowel $\rightarrow$ Vowel                   & 114   \\
        9 & Bilabial $\rightarrow$ Vowel               & 104   \\
        10 & Dental $\rightarrow$ Alveolar/Palatal      & 98    \\ \bottomrule
    \end{tabular}
    \caption*{The high frequency of confusions originating from Alveolar/Palatal phonemes (e.g., /S/, /Z/, /SH/, /ZH/, /T/, /D/) directly relates to their reliance on high-frequency spectral energy (4-8 kHz and above), which is most vulnerable to presbycusis. The most frequent substitution, Alveolar/Palatal $\rightarrow$ Labiodental (e.g., /S/ $\rightarrow$ /F/), exemplifies a shift towards lower-frequency fricatives when higher frequencies are attenuated. Intra-category confusions (Alveolar/Palatal $\rightarrow$ Alveolar/Palatal) likely reflect voicing or manner errors within high-frequency sounds, consistent with high-frequency cue degradation. Vowel-to-Vowel confusions (e.g., /IY1/ $\rightarrow$ /EY1/) indicate degradation of higher formants (F2, F3) crucial for vowel discrimination. The significant proportion of deletions further highlights the complete loss of information for high-frequency phonemes, often at word endings or in clusters, consistent with sounds falling below threshold.}
\end{table}

\begin{figure}[H]
    \centering
    \includegraphics[width=0.8\textwidth]{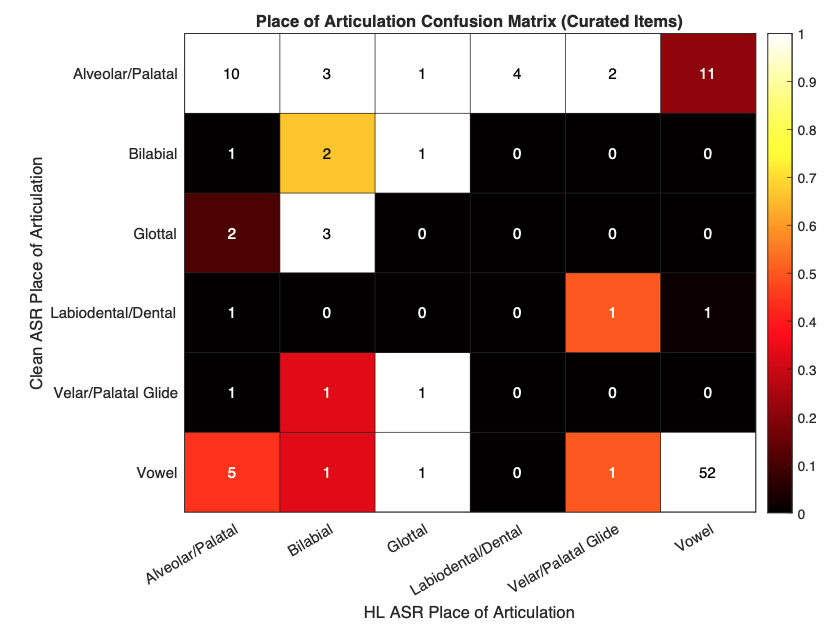} 
    \caption{Place of Articulation Confusion Matrix (Curated Items). This heatmap depicts the distribution of place of articulation confusions within the curated test item set, specifically for substitution errors. The rows represent the place of articulation of the phoneme in the Clean ASR output, and the columns represent the place of articulation of the confused phoneme in the HL ASR output. The color intensity within each cell indicates the normalized count of observed substitutions. This visualization confirms whether the curated items predominantly reflect confusions among phonemes with high-frequency acoustic cues (e.g., Alveolar/Palatal fricatives) and specific patterns of misidentification (e.g., shifts to Labiodental place), consistent with the effects of high-frequency hearing loss.}
    \label{fig:articulation_matrix}
\end{figure}

The high frequency of confusions originating from Alveolar/Palatal phonemes (e.g., /S/, /Z/, /SH/, /ZH/, /T/, /D/) directly relates to their reliance on high-frequency spectral energy (4-8 kHz and above), which is most vulnerable to presbycusis. The most frequent substitution, Alveolar/Palatal $\rightarrow$ Labiodental (e.g., /S/ $\rightarrow$ /F/), exemplifies a shift towards lower-frequency fricatives when higher frequencies are attenuated. Intra-category confusions (Alveolar/Palatal $\rightarrow$ Alveolar/Palatal) likely reflect voicing or manner errors within high-frequency sounds, consistent with high-frequency cue degradation. Vowel-to-Vowel confusions (e.g., /IY1/ $\rightarrow$ /EY1/) indicate degradation of higher formants (F2, F3) crucial for vowel discrimination. The significant proportion of deletions further highlights the complete loss of information for high-frequency phonemes, often at word endings or in clusters, consistent with sounds falling below threshold. This is further visualized in Figure \ref{fig:articulation_matrix}, which depicts the distribution of place of articulation confusions within the curated test item set for substitution errors, confirming that curated items predominantly reflect confusions among phonemes with high-frequency acoustic cues.

\subsection{Curated Test Item Characteristics}
\begin{figure}[H]
    \centering
    \includegraphics[width=\textwidth]{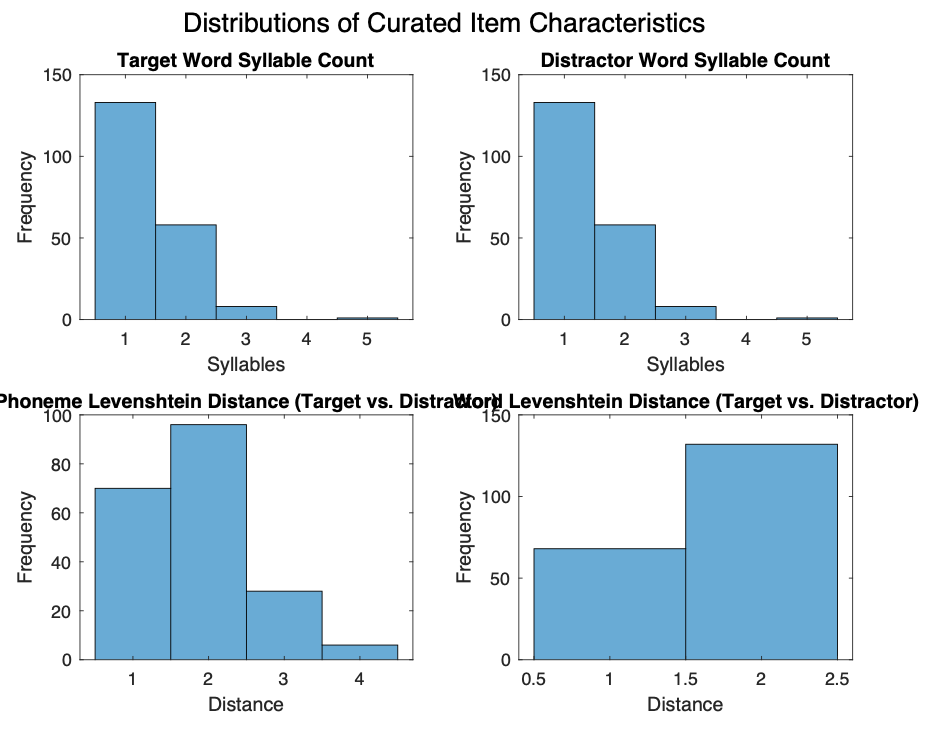} 
    \caption{Distributions of Curated Item Characteristics (Syllable and Levenshtein Distances). This panel of histograms illustrates: (a) target word syllable count, (b) distractor word syllable count, (c) phoneme Levenshtein distance (target vs. distractor), and (d) word Levenshtein distance (target vs. distractor). These figures confirm that criteria like matching syllable counts and limiting word Levenshtein distance were successfully applied, ensuring perceptual and lexical similarity.}
    \label{fig:item_characteristics}
\end{figure}

The curation process resulted in 200 diagnostically targeted word pairs, demonstrating the successful application of the filtering criteria. Figure \ref{fig:item_characteristics} illustrates characteristics of these curated items, including histograms of target and distractor word syllable counts, and phoneme and word Levenshtein distances, confirming that criteria like matching syllable counts and limiting word Levenshtein distance were successfully applied.

\begin{figure}[H]
    \centering
    \includegraphics[width=0.7\textwidth]{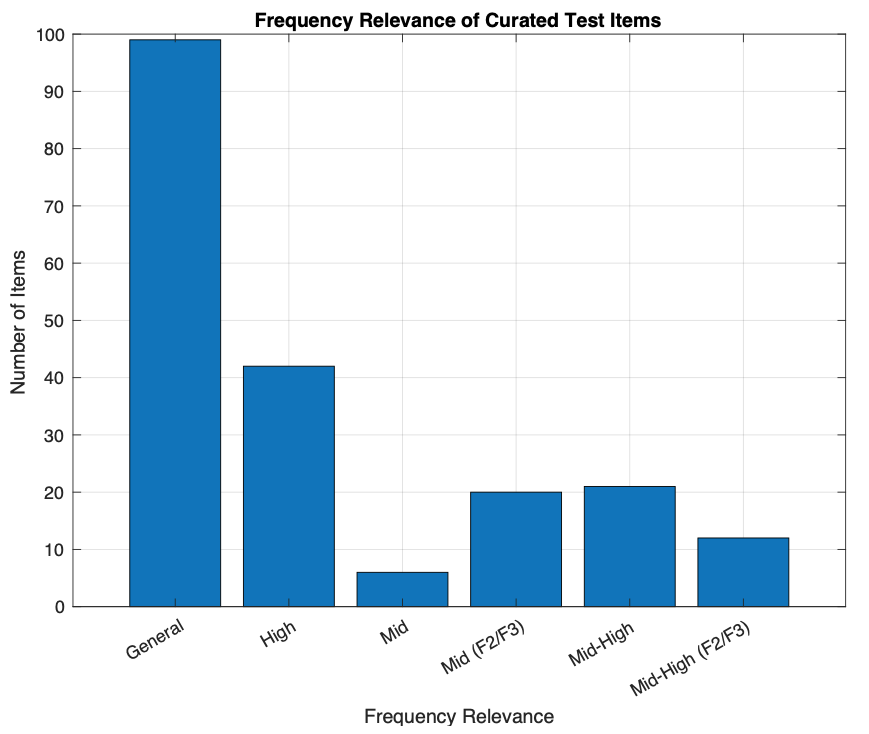} 
    \caption{Frequency Relevance of Curated Test Items. This bar chart shows the distribution of curated items based on the "Frequency Relevance" of the CleanPhonemeInvolved (e.g., 'High', 'Mid-High', 'Mid', 'General'). This quantifies the extent to which the test battery targets speech sounds whose perception is critically dependent on mid-to-high frequency information, aligning with the diagnostic objectives for presbycusis.}
    \label{fig:frequency_relevance}
\end{figure}

The distribution of curated items based on the "Frequency Relevance" of the CleanPhonemeInvolved (e.g., 'High', 'Mid-High', 'Mid', 'General') is shown in Figure \ref{fig:frequency_relevance}. This quantifies the extent to which the test battery targets speech sounds whose perception is critically dependent on mid-to-high frequency information, aligning with the diagnostic objectives for presbycusis.

\begin{figure}[H]
    \centering
    \includegraphics[width=0.7\textwidth]{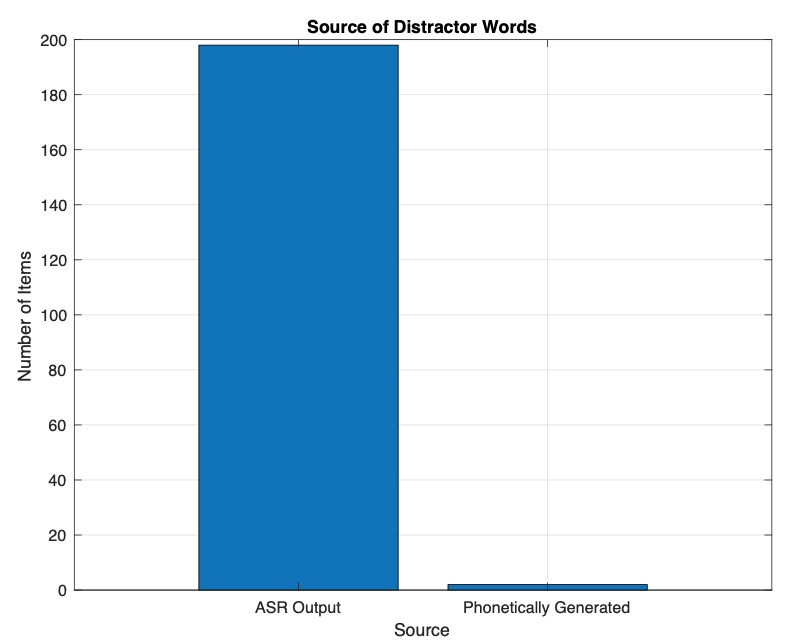} 
    \caption{Source of Distractor Words. This bar chart indicates the proportion of distractor words derived directly from ASR output under simulated HL versus those phonetically generated. This figure provides insight into the reliance on real-world ASR errors versus synthetically created plausible confusions in the final test battery.}
    \label{fig:distractor_source}
\end{figure}

Figure \ref{fig:distractor_source} indicates the proportion of distractor words derived directly from ASR output under simulated HL versus those phonetically generated, providing insight into the reliance on real-world ASR errors versus synthetically created plausible confusions in the final test battery.

The final selected word pairs for the test, along with their diagnostic performance metrics (NH Correct Percentage, HL Correct Percentage, and Diagnostic Difference), are provided in Table \ref{tab:diagnostic_items}. These exemplify the highest diagnostic differences between NH and HL ASR performance, crucial for the test's discriminative power.
\begin{longtable}{@{} >{\raggedright\arraybackslash}p{2.2cm} >{\raggedright\arraybackslash}p{2.2cm} c c c @{}}
    \caption{Selected Test Items with Diagnostic Performance Metrics.}
    \label{tab:diagnostic_items} \\
    \toprule
    OriginalWord & Distractor & NH\_Perc\_corr & HL\_Perc\_corr & Difference \\ 
    \midrule
    \endfirsthead 

    \multicolumn{5}{c}%
    {{\tablename\ \thetable{} -- continued from previous page}} \\
    \toprule
    OriginalWord & Distractor & NH\_Perc\_corr & HL\_Perc\_corr & Difference \\ 
    \midrule
    \endhead 

    \midrule
    \multicolumn{5}{r}{{Continued on next page}} \\
    \endfoot 

    \bottomrule
    \endlastfoot 

    object       & eject          & 100                   & 8                     & 92                    \\
    girls        & girl           & 88                    & 4                     & 84                    \\
    challenged   & challenge      & 80                    & 4                     & 76                    \\
    repainting   & recanting      & 98                    & 34                    & 64                    \\
    around       & 'round         & 96                    & 36                    & 60                    \\
    musical      & musica         & 84                    & 34                    & 50                    \\
    boys         & boyce          & 98                    & 54                    & 44                    \\
    even         & given          & 100                   & 60                    & 40                    \\
    effects      & effect         & 58                    & 18                    & 40                    \\
    few          & feel           & 80                    & 42                    & 38                    \\
    lost         & ast            & 98                    & 62                    & 36                    \\
    shall        & chalk          & 82                    & 46                    & 36                    \\
    wash         & wat            & 92                    & 58                    & 34                    \\
    even         & aven           & 98                    & 64                    & 34                    \\
    keep         & kip            & 94                    & 62                    & 32                    \\
    break        & bray           & 56                    & 26                    & 30                    \\
    morning      & earning        & 100                   & 70                    & 30                    \\
    suit         & said           & 60                    & 30                    & 30                    \\
    ability      & debility       & 46                    & 20                    & 26                    \\
    substances   & subspaces      & 34                    & 8                     & 26                    \\
    why          & whit           & 100                   & 74                    & 26                    \\
    this         & the            & 58                    & 32                    & 26                    \\
    employees    & employers      & 56                    & 32                    & 24                    \\
    cost         & aust           & 36                    & 12                    & 24                    \\
    dog          & god            & 60                    & 36                    & 24                    \\
    made         & mad            & 86                    & 64                    & 22                    \\
    every        & never          & 64                    & 44                    & 20                    \\
    hot          & pot            & 64                    & 44                    & 20                    \\
    makes        & mak            & 80                    & 60                    & 20                    \\
    nectar       & ector          & 92                    & 72                    & 20                    \\
    talked       & balke          & 84                    & 66                    & 18                    \\
    miles        & filed          & 100                   & 84                    & 16                    \\
    rag          & bagg           & 34                    & 18                    & 16                    \\
    year         & dear           & 66                    & 50                    & 16                    \\
    carry        & capri          & 68                    & 54                    & 14                    \\
    gone         & bode           & 94                    & 80                    & 14                    \\
    lunch        & blanch         & 98                    & 86                    & 12                    \\
    ahead        & behead         & 70                    & 60                    & 10                    \\
    dark         & barg           & 100                   & 90                    & 10                    \\
    his          & ein            & 100                   & 90                    & 10                    \\
    most         & move           & 46                    & 36                    & 10                    \\
    often        & bolten         & 94                    & 84                    & 10                    \\
    brother      & bother         & 48                    & 40                    & 8                     \\
    conviction   & convictions    & 100                   & 92                    & 8                     \\
    hands        & hand           & 98                    & 90                    & 8                     \\
    please       & cleave         & 100                   & 92                    & 8                     \\
    shredded     & threaded       & 76                    & 68                    & 8                     \\
    water        & beater         & 92                    & 84                    & 8                     \\
    beans        & bains          & 32                    & 26                    & 6                     \\
    dark         & duck           & 16                    & 10                    & 6                     \\
    wire         & ire            & 98                    & 92                    & 6                     \\ 
\end{longtable}

\subsection{Simulated Test Administration Results}
The simulation of test administration with 50 Normal Hearing and 50 Hearing Impaired participants, each assessed on a random subset of 50 items, yielded preliminary diagnostic performance metrics. The simulation parameters included a base moderate sloping HL audiogram for impaired participants with $\pm$10 dB variability, and a psychometric function incorporating phoneme Levenshtein distance with a sigmoid slope (k=0.8) and a HL impact factor (0.05 per dB HL). A test failure threshold of 80\% correct was used.

\begin{figure}[H]
    \centering
    \includegraphics[width=\textwidth]{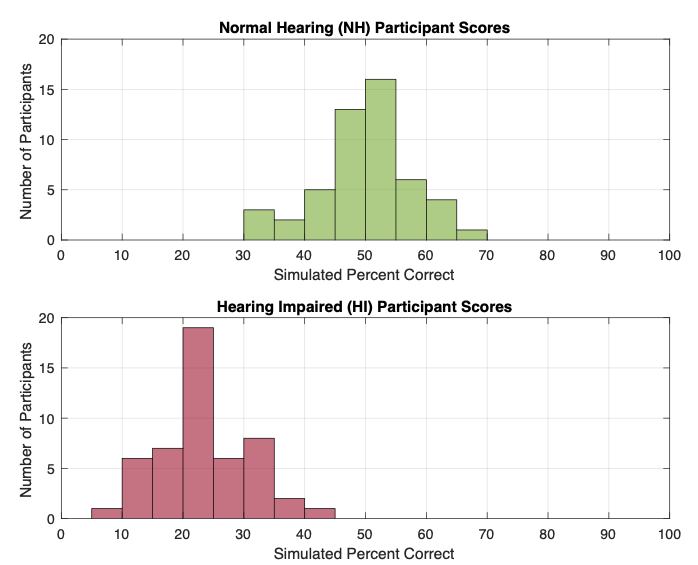} 
    \caption{Simulated Percent Correct Histograms by Participant Group. This figure displays two histograms, showing the distribution of simulated percent correct scores for both Normal Hearing (NH) and Hearing Impaired (HI) participant groups. This figure provides insight into the spread and central tendency of performance within each group.}
    \label{fig:score_histograms}
\end{figure}

The distribution of simulated percent correct scores for both Normal Hearing (NH) and Hearing Impaired (HI) participant groups is shown in Figure \ref{fig:score_histograms}, providing insight into the spread and central tendency of performance within each group.

\begin{figure}[H]
    \centering
    \includegraphics[width=0.7\textwidth]{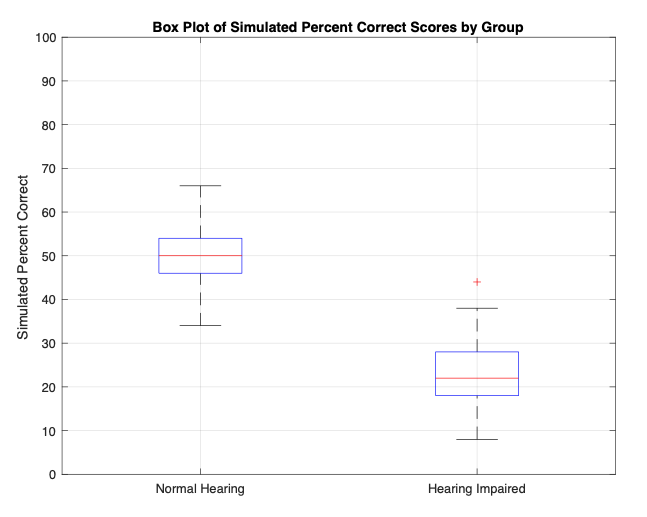} 
    \caption{Box Plot of Simulated Percent Correct Scores by Group. This box plot visually represents the distribution of simulated percent correct scores for Normal Hearing and Hearing Impaired participant groups. The central mark indicates the median, the box edges represent the 25th and 75th percentiles, and the whiskers extend to the most extreme data points not considered outliers.}
    \label{fig:score_boxplot}
\end{figure}

Figure \ref{fig:score_boxplot} further visualizes this distribution using a box plot.

\begin{figure}[H]
    \centering
    \includegraphics[width=0.7\textwidth]{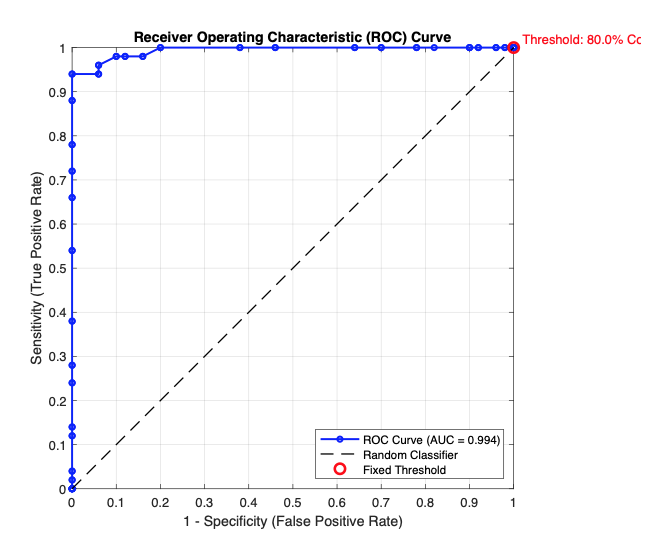} 
    \caption{Receiver Operating Characteristic (ROC) Curve. This figure presents the ROC curve, illustrating the trade-off between Sensitivity and (1 - Specificity) across various potential test failure thresholds. The dashed black line represents the performance of a random classifier. A marker indicates the performance at the chosen fixed test failure threshold of 80\% correct. The Area Under the Curve (AUC) indicates the overall discriminative power of the test.}
    \label{fig:roc_curve}
\end{figure}

The Receiver Operating Characteristic (ROC) curve, illustrating the trade-off between Sensitivity and (1 - Specificity) across various potential test failure thresholds, is presented in Figure \ref{fig:roc_curve}. The Area Under the Curve (AUC) indicates the overall discriminative power of the test.

\begin{figure}[H]
    \centering
    \includegraphics[width=0.6\textwidth]{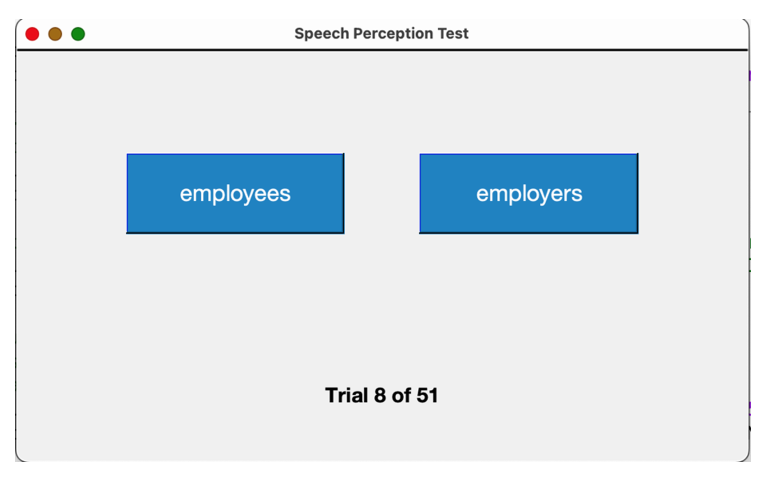} 
    \caption{Interface for the speech test so that a human can do it. This figure shows the user interface designed for the speech perception test, allowing a human participant to interact with the test by selecting one of two presented word options. It typically displays two buttons with the word choices and a trial counter.}
    \label{fig:test_interface}
\end{figure}

Finally, Figure \ref{fig:test_interface} shows an example of the interface designed for the human speech test.

\clearpage 
\section{Discussion}
This study aimed to develop a novel ASR-based frequency-specific speech test to overcome the limitations of traditional audiometry in characterizing the functional impact of hearing loss, particularly presbycusis. By simulating hearing loss effects on ASR performance, the goal was to derive granular, phoneme-level insights into speech perception deficits. The findings demonstrate the feasibility of this approach and lay a foundation for a more diagnostically informative hearing assessment tool.
In response to our first research question, the methodology successfully utilized an ASR system to simulate the perceptual effects of hearing loss and generate comprehensive phoneme confusion profiles. By processing speech stimuli under controlled acoustic degradation, mimicking typical hearing loss conditions, the ASR system provided detailed error patterns. This approach inherently leverages the ASR's capacity for phoneme-level scoring, offering a more granular understanding of perception errors than traditional whole-word or sentence scoring \cite{hnath2016}. The observed shifts in ASR outputs under simulated hearing loss serve as a quantitative proxy for how specific speech sound cues might be distorted or lost to a human listener.
Addressing the second research question, the specific phoneme-level confusion patterns observed strongly correlate with the acoustic properties of speech sounds and known effects of presbycusis. The high prevalence of substitutions and deletions, particularly involving high-frequency phonemes like sibilants and alveolar/palatal stops, is consistent with the spectral shaping characteristic of sloping high-frequency hearing loss. For instance, the most frequent substitution of alveolar/palatal phonemes for labiodental ones (e.g., /S/ to /F/, see Table \ref{tab:phoneme_confusions} and Figure \ref{fig:articulation_matrix}) directly reflects the attenuation of high-frequency energy essential for discriminating sibilants, which pushes recognition towards sounds with lower peak frequencies. Similarly, vowel-to-vowel confusions (e.g., /IY1/ to /EY1/) suggest the degradation of higher formants (F2, F3) crucial for vowel distinctiveness, a known challenge in presbycusis. The significant proportion of deletions further indicates a complete loss of information for certain phonemes, often high-frequency consonants that fall below threshold (see Figure \ref{fig:error_distribution}). These patterns align with psychoacoustic understanding of how hearing loss impacts the perception of speech cues. Importantly, because our simulation primarily models frequency-dependent attenuation, its findings are particularly pertinent to mild-to-moderate hearing loss, where *audibility* is considered a dominant factor limiting speech perception, especially in quiet \cite{Plomp1978}. Our ASR-derived confusions (e.g., /S/→/F/, high-frequency deletions) directly reflect this loss of audible cues \cite{Giguere1995}.
Regarding the third research question, the simulation demonstrated that a test battery curated from these ASR-derived confusions can effectively differentiate between normal-hearing and hearing-impaired listeners in a simulated environment. The diagnostic difference calculated for curated word pairs indicated their discriminative power (see Table \ref{tab:diagnostic_items}). The simulated test administration, as shown by distinct distributions of percent correct scores (Figure \ref{fig:score_histograms} and Figure \ref{fig:score_boxplot}) and a quantifiable ROC curve (Figure \ref{fig:roc_curve}), confirmed the test's potential to distinguish between the two groups. This provides a strong preliminary indication of the test's diagnostic utility in a controlled setting.
Finally, addressing the critical considerations for translating this ASR-based diagnostic test into a practical tool for human participants (RQ4), several factors are paramount. Firstly, while ASR serves as an effective proxy for large-scale simulation and identifying robust confusion patterns, it does not perfectly replicate the complex, non-linear auditory processing of human listeners, especially supra-threshold deficits like reduced frequency resolution or impaired temporal processing that are often central to human hearing loss \cite{Oxenham2008, Henry2022, mdpi2025}. The current simulation relies on acoustic filtering, which is a simplification of the multi-faceted nature of sensorineural hearing loss. Therefore, direct validation with human participants is an indispensable next step. The diversity of speech patterns (accents, disordered speech) and environmental noise conditions in real-world clinical settings also necessitates robust ASR performance beyond controlled laboratory conditions \cite{gutz2025}.
A key aspect for real-world application is the calibration of a free parameter: the level of added noise. It is generally understood that ASR systems often perform worse than human listeners (both normal hearing and those with hearing loss) in noisy conditions, particularly when dealing with unpredictable noise types or lower signal-to-noise ratios \cite{Cooke2006}. By strategically adding a certain level of noise to the stimuli presented to the ASR, we can potentially degrade its performance to a level that more closely mirrors that of human listeners under similar conditions. This calibration is crucial for ensuring the ASR-based test provides a realistic and comparable assessment. However, the optimal level of this added noise is currently unknown and will require careful calibration with data from human participants with varying hearing abilities, across a range of SNRs, to establish a reliable mapping between ASR and human performance.
Despite these considerations, the ASR-based approach offers several advantages. It enables the rapid development and objective scoring of speech tests, potentially reducing clinician time and standardizing assessment procedures \cite{mdpi2025, polspoel2025}. The granular phoneme-level insights gained could lead to more personalized diagnostic reports that go beyond a simple audibility measure, offering a "confusion profile" that maps specific perceptual difficulties to affected frequency regions.
\subsection*{Limitations}
The primary limitation is the reliance on ASR as a proxy for human listeners. While ASR models are increasingly sophisticated, their internal processing mechanisms may not fully align with human auditory perception, particularly regarding specific forms of auditory distortion inherent to hearing loss. The current simulation of hearing loss is purely acoustic (filtering and noise) and does not encompass more complex physiological aspects of cochlear damage or central auditory processing deficits. While this aligns with the *audibility*-centric view of mild-to-moderate loss \cite{Plomp1978, Humes2007}, future work should consider incorporating models of supra-threshold deficits \cite{Oxenham2008, Henry2022} for broader applicability, especially in noise. The TIMIT corpus, while phonetically rich, consists of read speech and may not fully represent the variability of natural conversational speech.
\subsection*{Future Work}
The immediate next step involves pilot testing the curated test battery with human participants, including normal-hearing individuals and those with presbycusis, to validate the ASR-derived confusions and diagnostic predictions in a real-world setting. Future refinements to the methodology should include exploring more sophisticated phonetic alignment algorithms for a more complete analysis of insertions and deletions in confusion matrices. Further research into adaptive or personalized distractor selection strategies, potentially involving phonetically motivated minimal pairs beyond direct ASR confusions, could enhance test specificity. Development of an automated diagnostic report generation tool that visualizes the "Confusion Profile" or an "Audiogram of Confusion" would greatly enhance clinical utility. Finally, investigating the integration of this framework with advanced AI models such as Speech Foundation Models and Large Audio-Language Models may offer even more nuanced speech intelligibility prediction and dynamic stimulus generation capabilities, potentially leading to more personalized and precise hearing assessments \cite{zhou2025, yang2025}. Addressing the practical challenges of clinical workflow integration and ensuring cost-effectiveness will be crucial for the widespread adoption of such innovative diagnostic tools \cite{reed_nd}.

\end{document}